\documentclass[prb,aps,amsmath,amssymb,amsfonts,floatfix,superscriptaddress,showpacs,twocolumn]{revtex4}
\usepackage{graphicx}
\usepackage{color}
\usepackage[utf8]{inputenc}
\usepackage{dcolumn}
\usepackage{hyperref}
\usepackage[caption=false]{subfig}
\usepackage[english]{babel}
\usepackage{indentfirst}
\hypersetup{colorlinks=true,breaklinks,linkcolor=blue,urlcolor=blue,citecolor=blue}

\usepackage{tikz}
\usetikzlibrary{calc}
\usetikzlibrary{fit}
\usetikzlibrary{decorations.pathmorphing}
\usetikzlibrary{decorations.pathreplacing}

\renewcommand{\i}{{\ensuremath{\imath}}}
\newcommand\Let{\mathrel{\mathop:\!\!=}}

\newcommand{\inu}{{\ensuremath{\i\nu}}}
\newcommand{\kv}{\ensuremath{\mathbf{k}}}
\newcommand{\KV}{\ensuremath{\mathbf{K}}}

\newcommand{\qv}{\ensuremath{\mathbf{q}}}

\newcommand{\av}[1]{\ensuremath{\left\langle #1 \right\rangle}}

\newcommand{\cdag}{{\ensuremath{\mathop{c^{\dagger}}}}}

\def \Re {\mathop {\rm Re}}
\def \Tr {\mathop {\rm Tr}}

\newcommand\varpm{\mathbin{\vcenter{\hbox{%
  \oalign{\hfil$\scriptstyle+$\hfil\cr
          \noalign{\kern-.3ex}
          $\scriptscriptstyle({-})$\cr}%
}}}}
\newcommand\varmp{\mathbin{\vcenter{\hbox{%
  \oalign{$\scriptstyle({+})$\cr
          \noalign{\kern-.3ex}
          \hfil$\scriptscriptstyle-$\hfil\cr}%
}}}}

\begin{document}

\title{Second-order dual fermion approach to the Mott transition\\ in the two-dimensional Hubbard model}

\author{Erik G. C. P. van Loon}
\affiliation{Radboud University, Institute for Molecules and Materials, NL-6525 AJ Nijmegen, The Netherlands}

\author{Mikhail I. Katsnelson}
\affiliation{Radboud University, Institute for Molecules and Materials, NL-6525 AJ Nijmegen, The Netherlands}

\author{Hartmut Hafermann}
\affiliation{Mathematical and Algorithmic Sciences Lab, Paris Research Center, Huawei Technologies France SASU, 92100 Boulogne Billancourt, France}

\begin{abstract}
We apply the dual fermion approach with a second-order approximation to the self-energy to the Mott transition in the two-dimensional Hubbard model.
The approximation captures nonlocal dynamical short-range correlations as well as several features observed in studies using cluster dynamical mean-field theory. This includes a strong reduction of the critical interaction and inversion of the slope of the transition lines with respect to single-site dynamical mean-field theory. 
We show that these effects coincide with a much smaller momentum differentiation compared to cluster methods. We further discuss the role of the self-consistency condition and show that the approximation behaves as an asymptotic series at low temperature.
\end{abstract}


\maketitle

\clearpage

\section{Introduction}

%Mott MIT, insights from DMFT and cluster methods
The Mott metal-insulator transition (MIT) is a fundamental problem in condensed matter physics. Dynamical mean-field theory (DMFT) has provided important insights, because the low-energy coherent quasiparticle excitations as well as the incoherent high-energy excitations can be treated on the same footing~\cite{Georges96}.
The transition was later analyzed using cluster extensions of DMFT. Cellular DMFT results~\cite{Zhang07,Park08} revealed a substantial reduction of the critical interaction and a reversal of the slope of the transition lines. Subsequent DCA studies on larger clusters found a similar reduction of the critical interaction, but mainly focused on a different aspect of the transition~\cite{Ferrero09,Werner09,Gull09}: For both the interaction-driven and doping-driven transition, an intermediate sector-selective phase was found, in which some momentum sectors become insulating while others remain metallic. In the studies on larger clusters ($\geq 8$ sites) the momentum resolution was sufficiently high to clearly separate the nodal and antinodal directions. All of them consistently showed that quasiparticles are first destroyed in the antinodal direction, leading to an 'orbital-selective transition in momentum space'.

In cluster methods, the self-energy is treated rigorously, but correlations are restricted to the size of the cluster. Even though the cluster size acts as a control parameter, convergence is seldom reached in practice due to the exponential growth in computational complexity. Diagrammatic extensions of DMFT~\cite{Rohringer18}, like the dynamical vertex approximation (D$\Gamma$A)~\cite{Toschi07} and dual fermion (DF)~\cite{Rubtsov08}, provide a complimentary viewpoint.
In these approaches, nonlocal corrections to the DMFT self-energy are included diagrammatically. 
The local interaction vertices are given by (reducible or irreducible) vertex functions of an impurity model.
A significantly higher momentum resolution can be reached, while the self-energy in practice remains approximate at any scale.
In general the results necessarily depend on the diagrammatic approximation.
For small systems, all diagrammatic corrections obtained from a two-particle vertex can be obtained 
by solving the full set of parquet equations~\cite{Valli2015}. Another means to avoid a possible bias is to sample diagrams irrespective of their topology using a diagrammatic Monte Carlo approach~\cite{Iskakov2016,Gukelberger2017}.
When diagrams in a given fluctuation channel can be expected to dominate for physical reasons, the ladder approximation is a more sophisticated alternative. Examples are the ladder dynamical vertex approximation  (D$\Gamma$A)~\cite{Toschi07} or the ladder DF approximation (LDFA)~\cite{Hafermann09}. These approximation can even yield quantitatively accurate results in regimes where they capture the dominant fluctuations~\cite{Hafermann09,Gukelberger2017}.
Recent systematic studies based on these approaches have revealed precursors of the Mott insulator at low temperature and interaction strengths significantly smaller than the critical value obtained in CDMFT~\cite{Schafer15,vanLoon17}.

The summation of the infinite diagram series in the ladder approximation can be computationally demanding for inhomogeneous systems~\cite{Takemori16,Takemori18} or multiorbital systems. In addition it requires a regularization procedure if the series initially diverges in the dynamical-mean solution~\cite{Otsuki2014}, whose computational complexity typically increases with the strength of the fluctuations.
A much simpler approximation is the second-order dual fermion approximation~\cite{Rubtsov08,Rubtsov09}, which we refer to as DF$^{(2)}$.
The approximation has been shown to capture short-range spatial correlations beyond DMFT.
So far, a systematic study based on the DF$^{(2)}$ approximation and the role of the self-consistency condition has been lacking. For applications it is useful to understand its physical content and limitations.
In this paper we aim to close this gap by reexamining the Mott transition in the two-dimensional Hubbard model within DF$^{(2)}$ as a benchmark. We also report results obtained with an alternative self-consistency condition.

\section{Model}
\label{sec:model}

We apply our method to the single-band Hubbard model on the two-dimensional square lattice
\begin{align}
\label{H}
H=\sum_{\kv \sigma}\epsilon_{\kv}c_{\kv\sigma}^{\dagger}c_{\kv\sigma} + U\sum_{i}n_{i\uparrow}n_{i\downarrow}.
\end{align}
The noninteracting dispersion
\begin{align}
\epsilon_{\kv}=-2t(\cos k_{x}+\cos k_{y}) - 4t'(\cos k_{x}\cos k_{y}),
\end{align}
is determined by nearest- ($t$) and next-nearest neighbor hopping $t'$. We take $t=1$ as the energy unit. In the above, latin indices denote lattice sites, $\sigma=\uparrow,\downarrow$ labels spin, $n_{i\sigma}=\cdag_{i\sigma}c_{i\sigma}$ and $U$ is the local Hubbard repulsion.

In particular the two-dimensional model is well-studied and often used as a benchmark for new computational approaches~\cite{LeBlanc15}. An exact solution to the interacting Hubbard model is known in the limit of infinite dimensions~\cite{Metzner89} provided by DMFT~\cite{Georges96}. The DMFT solutions has provided important insights into the Mott metal-insulator transition.
Cluster extensions of DMFT~\cite{Maier05} treat short-range correlations exactly and refine this picture: In cellular DMFT the critical interaction is considerably reduced due to the effect of spatial correlations~\cite{Zhang07,Park08}. The dynamical cluster approximation indicates a strong momentum space differentiation and a momentum sector-selective opening of the gap. Below we address how these aspects are reflected in the DF$^{(2)}$ approximation.
Diagrammatic extensions~\cite{Rohringer18} provide a complementary viewpoint, because they can treat also long-range spatial correlations, albeit perturbatively.

\section{Method}
\label{sec:method}

In the following, we briefly sketch the derivation of  the dual fermion method to provide some intuition and to define all necessary quantities. A modern introduction and derivation can be found in review of Ref.~\onlinecite{Rohringer18}.
The dual fermion approach is an action-based method. The basic idea is to introduce an auxiliary quantum impurity model at each lattice site (we assume translational invariance in the following). This may be done leaving the original action unaltered by formally adding and subtracting a hybridization function at each site. The result can be written in the form
\begin{align}
\label{slat_rew}
S_{\text{lat}}=\sum_{i}S_{\text{imp}}[c_{i}^{*},c_{i}] &- \sum_{\kv\nu\sigma}c^{*}_{\kv\nu\sigma}(\Delta_{\nu\sigma}-\varepsilon_{\kv})c_{\kv\nu\sigma},
\end{align}
where the impurity action $S_{\text{imp}}$ is given by
\begin{align}
\label{simp}
S_{\text{imp}}[c^{*},c]=&-\!\!\sum_{\nu\sigma} c^{*}_{\nu\sigma}[\inu+\mu-\Delta_{\nu\sigma}]c_{\nu\sigma}+ U\sum_{\omega}n_{\omega\uparrow}n_{-\omega\downarrow}.
\end{align}
Here $\nu$ ($\omega$) denote fermionic (bosonic) Matsubara frequencies. 

In order to formulate a Feynman-type diagrammatic expansion around the non-Gaussian impurity model, the second term in \eqref{slat_rew} is decoupled by introducing auxiliary (dual) degrees of freedom through a Hubbard-Stratonovich transformation. It is important that these couple locally to the original fermions, for two reasons: Firstly, this guarantees that the mapping to dual fermions preserves the topology of the diagrams. For example, a dual self-energy diagram connecting two neighboring sites will also introduce nearest-neighbor correlations in terms of the real fermions.
And secondly, the original fermions can be integrated out for each site separately. This produces the connected correlation functions of the impurity coupled to dual fermions. The latter enter the resulting dual action, which reads
\begin{align}
\tilde{S} = -\sum_{\kv\nu\sigma} f_{\kv\nu\sigma}^{*}(\tilde{G}^{0}_{\kv\nu\sigma})^{-1} f_{\kv\nu\sigma} + V[f^{*},f].
\end{align}
Here the bare dual Green's function is defined as
\begin{align}
\label{gd0}
\tilde{G}^{0}_{\kv\nu\sigma} = G_{\kv\nu\sigma}^{\text{DMFT}} - g_{\nu\sigma},
\end{align}
where $g_{\nu\sigma} \Let -\av{c_{\nu\sigma} c^{*}_{\nu\sigma}}$ denotes the interacting impurity Green's function. The interacting DMFT lattice Green's function is given by
\begin{align}
(G_{\kv\nu\sigma}^{\text{DMFT}})^{-1} = \inu +\mu -\epsilon_{\kv} - \Sigma_{\nu}
\label{eq:gDMFT}
\end{align}
and, to leading order, the dual interaction $V$ is
\begin{align}
\label{V}
V[f^{*},f] = - & \frac{1}{4} \sum_{\nu\nu'\omega}\sum_{\sigma_{i}}\gamma^{\sigma_{1}\sigma_{2}\sigma_{3}\sigma_{4}}_{\nu\nu'\omega}f_{\nu\sigma_{1}}^{*}f_{\nu+\omega,\sigma_{2}}f^{*}_{\nu'+\omega,\sigma_{3}}f_{\nu'\sigma_{4}}.
\end{align}
The reducible impurity vertex hence plays the role of the bare interaction of the dual fermions. It is defined as
\begin{align}
\label{gammadef}
\gamma_{\nu\nu'\omega}^{\sigma\sigma'} & \Let \frac{g^{\sigma\sigma'}_{\nu\nu'\omega} - \beta g_{\nu\sigma}g_{\nu'\sigma'}\delta_{\omega} + \beta g_{\nu\sigma}g_{\nu+\omega\sigma}\delta_{\nu\nu'}\delta_{\sigma\sigma'}
}{g_{\nu\sigma}g_{\nu+\omega,\sigma}g_{\nu'+\omega\sigma'}g_{\nu'\sigma'}}.
\end{align}
which involves the two-particle Green's function
\begin{align}
\label{chi4def}
g_{\nu\nu'\omega}^{\sigma\sigma'} & \Let \av{c_{\nu\sigma}c^{*}_{\nu+\omega,\sigma}c_{\nu'+\omega,\sigma'}c^{*}_{\nu'\sigma'}}.
\end{align}
In the above, we have introduced the short-hand notation $\gamma^{\sigma\sigma'}\Let \gamma^{\sigma\sigma\sigma'\sigma'}$.

\subsection{Dual self-energy}

\begin{figure}[h]
\includegraphics[scale=0.25,angle=0]{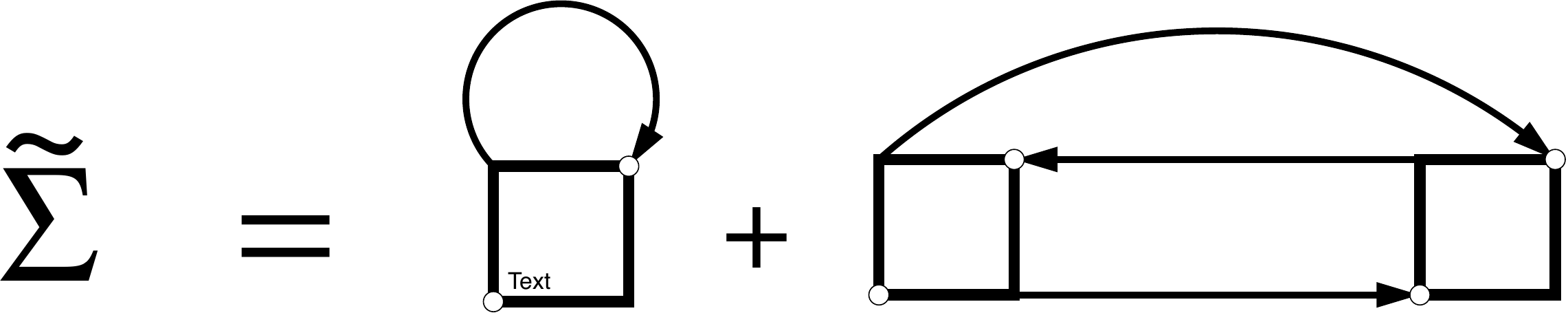} 
\caption{\label{fig:2ndorder} Second-order approximation to the dual self-energy.
The vertex function $\gamma$ is depicted by a square. Lines are fully dressed $\tilde{G}$ propagators.}
\end{figure}

The second-order approximation to the dual self-energy is depicted diagrammatically in Fig.~\ref{fig:2ndorder}.
In the paramagnetic case, it is convenient to work with the charge (c) and spin (s) components defined by $\gamma^{\text{s(c)}}=\gamma^{\uparrow\uparrow}\varmp\gamma^{\uparrow\downarrow}$.
Introducing
\begin{align}
\label{chid}
\tilde{\chi}^{0}_{\qv\omega\nu} = \frac{1}{N}\sum_{\kv'}\tilde{G}_{\kv'+\qv \nu+\omega}\tilde{G}_{\kv'\nu},
\end{align}
the dual self-energy in the DF$^{(2)}$ approximation reads
\begin{align}
\label{sigmad2}
\tilde{\Sigma}_{\kv\nu\sigma} =& -\frac{T}{N}\sum_{\kv'\nu'}\gamma_{\nu\nu'\omega=0}^{\text{c}}\tilde{G}_{\kv'\nu'} \notag\\
&-\frac{1}{2}\frac{T^{2}}{N}\sum_{\qv\omega}\sum_{\nu'}\Bigg[\frac{1}{2}\gamma_{\nu\nu'\omega}^{\text{c}}\tilde{G}_{\kv+\qv\nu+\omega}\tilde{\chi}^{0}_{\qv\omega\nu'}\gamma_{\nu'\nu\omega}^{\text{c}}\notag\\
&\qquad\qquad\qquad\ +\frac{3}{2}\gamma_{\nu\nu'\omega}^{\text{s}}\tilde{G}_{\kv+\qv\nu+\omega}
\tilde{\chi}^{0}_{\qv\omega\nu'}\gamma_{\nu'\nu\omega}^{\text{s}}\Bigg].
\end{align}
The physical self-energy is obtained from its dual counterpart through the relation
\begin{align}
\Sigma_{\kv\nu} = \Sigma_{\nu}^{\text{imp}} + \frac{\tilde{\Sigma}_{\kv\nu}}{1 + \tilde{\Sigma}_{\kv\nu}g_{\nu}}.
\label{eq:sigma}
\end{align}
Equivalently, one may obtain the physical Green's function $G_{\kv\nu}$ from the renormalized to dual Green's function $\tilde{G}_{\kv\nu}$ through the expression
\begin{align}
G_{\kv\nu} &\!\!=\! (\Delta_{\nu}\!-\!\epsilon_{\kv})^{-1}\!\! +\! (\Delta_{\nu}\!-\!\epsilon_{\kv})^{-1}g_{\nu}^{-1}\tilde{G}_{\kv\nu}g_{\nu}^{-1}(\Delta_{\nu}\!-\!\epsilon_{\kv})^{-1}.
\end{align}

\subsection{Self-consistency condition and computational scheme}

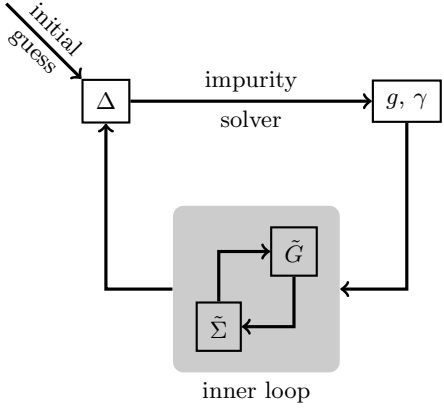
\begin{figure}
\begin{tikzpicture}
[
box/.style = {draw=black,thick,inner sep=5pt,minimum size=0mm,align=center},
abovebox/.style = {above}
]
 \coordinate (hybridization) at (0,0) ;
 \coordinate (impurity) at ($(hybridization)+(4,0)$) ;
 \coordinate (dualpt) at ($(hybridization)+(2.5,-2)$) ;
 \coordinate (dualsigma) at ($(dualpt)+(-1,-1.)$) ;

  \node [fit={($(dualpt)+(0.5,0.5)$) ($(dualsigma)+(-0.5,-0.5)$)},fill=gray!40,rounded corners] (graysq) {}  ;
 
 \node (hybridization_box) at (hybridization) [box] {$\Delta$} ;
 \node (impurity_box) at (impurity) [box] {$g$, $\gamma$} ;
 \node (dualpt_box) at (dualpt) [box] {$\tilde{G}$} ; 
 \node (dualsigma_box) at (dualsigma) [box] {$\tilde{\Sigma}$} ; 
 
 \draw[->,very thick] ($(hybridization_box.north west)+(-1,1)$) to node[midway,sloped,above]{initial} node[midway,sloped,below]{guess} ($(hybridization_box.north west)$) ;
  
 \draw[->,very thick] (hybridization_box.east) to node[midway,above]{impurity} node[midway,below]{solver} (impurity_box.west) ;
 
 \draw[->,very thick] ($(dualsigma_box.north)$) |- ($(dualpt_box.west)$) ;
 \draw[->,very thick] ($(dualpt_box.south)$) |- ($(dualsigma_box.east)$) ;
 
 \draw[->,very thick] (impurity_box.south) |- (graysq.east) ;
 \draw[->,very thick] (graysq.west) -| (hybridization_box.south) ;
 
 \node[below] at (graysq.south) {inner loop};
 
 \end{tikzpicture}
 \caption{
 Visualization of the computational scheme.
 }
 \label{fig:scheme}
\end{figure}

A priori, the choice of hybridization function is arbitrary. It was introduced such that in an exact theory, the solution is independent of its value. In practice however, it acts to improve the starting point of the perturbation theory.
There are two apparent self-consistency conditions to fix the hybridization function. 
Condition (i) has been used exclusively in the literature so far. It is obtained by requiring the local part of the \emph{interacting} dual Green's function to vanish:
\begin{align}
\text{SC (i): }&&
\frac{1}{N}\sum_{\kv}\tilde{G}_{\kv\nu} &= 0.
\label{eq:sc1}
\end{align}
In the following, we refer to this condition as the dual self-consistency condition. Condition (ii) is familiar from DMFT and reads
\begin{align}
\text{SC (ii): }&&
g_{\nu} &= \frac{1}{N}\sum_{\kv}G_{\kv\nu}.
\label{eq:sc2}
\end{align}
We refer to it as the lattice condition.
When no dual self-energy diagrams are taken into account, this condition is seen to be equivalent to requiring the local part of the bare dual Green function to be zero, $(1/N)\sum_{\kv}\tilde{G}_{\kv\nu}^{0} = 0$, by virtue of Eqs.~\eqref{gd0} and~\eqref{eq:gDMFT}. This shows that a theory of noninteracting dual fermions is equivalent to DMFT~\cite{Rubtsov08}. As soon as dual self-energy diagrams are taken into account however, this no longer holds and the resulting hybridization will be different from the DMFT. The physical lattice Green function then contains a momentum dependent self-energy, Eq.~\eqref{eq:sigma}.

The two conditions are not equivalent in general, except in case of noninteracting dual fermions, where they both correspond to DMFT. We discuss the physical content of these conditions in the results section below.

In practice, we enforce the self-consistency condition through an iterative update of the hybridization function, similar to DMFT.
For example, condition (i) is enforced by the update rule
\begin{align}
\label{update2}
\text{SC (i): }&&
\Delta_{\nu}^{\text{new}} &= \Delta_{\nu}^{\text{old}} + \xi[g_{\nu}^{-1}\tilde{G}_{\nu}^{\text{loc}}(G_{\nu}^{\text{loc}})^{-1}],
\end{align}
where $\xi\in(0,1]$ is a mixing parameter. Convergence is reached when $\tilde{G}_{\nu}^{\text{loc}}=0$. The other Green's functions only act as scaling factors. They are chosen such that for $\tilde{G}=\tilde{G^{0}}$, the angular bracket evaluates to $g_{\nu}^{-1}-(G_{\nu}^{\text{loc,DMFT}})^{-1}$.
Condition (ii) is enforced through the update
\begin{align}
\label{update}
\text{SC (ii): }&&
\Delta_{\nu}^{\text{new}} &= \Delta_{\nu}^{\text{old}} + \xi[g_{\nu}^{-1}-(G_{\nu}^{\text{loc}})^{-1}].
\end{align}

In summary, the computational scheme is as follows:
\begin{itemize}
\item[0.] Generate an initial guess for the hybridization function $\Delta$ (e.g. from DMFT).
\item[1.] For a given $\Delta$ compute $g$, $\Sigma_\text{imp}$ and $\gamma$.
\item[2.] Evaluate the diagrams, Eq.~\eqref{sigmad2}, to compute $\tilde{\Sigma}$.
\item[3.] Compute an updated $\tilde{G}$ from the dual Dyson equation, $\tilde{G}^{-1} = \tilde{G}^{-1}_{0} - \tilde{\Sigma}$.
\item[4.] Repeat steps 2.--3. until convergence (inner loop).
\item[5.] From a converged $\tilde{G}$ compute an update of the hybridization function $\Delta$ according to \eqref{update2}.
\item[6.] Repeat steps 1.--5. until convergence (outer loop).
\end{itemize}

This procedure is illustrated in Fig.~\ref{fig:scheme}.
If steps 2. to 4. are omitted (inner loop), the outer loop is equivalent to DMFT. In this case $\gamma$ does not have to be calculated. The computational effort for a dual fermion iteration is dominated by the measurement of the vertex function. 

\section{Results}

The following results are obtained for a lattice of size $64\times 64$ sites as a compromise between fine momentum resolution and consumption of computational and memory resources.
For the solution of the impurity model, we use the implementation of the segment picture variant of the continuous-time hybridization expansion algorithm\cite{Werner06} of Ref.~\onlinecite{Hafermann13} with improved estimators for the self-energy and vertex function\cite{Hafermann12}.
Here we employ the self-consistency condition (i). We compare the two alternative conditions in Sec.~\ref{sec:sc}.

\subsection{Mott transition}
\label{sec:Mott}

\begin{figure}[t]
\includegraphics{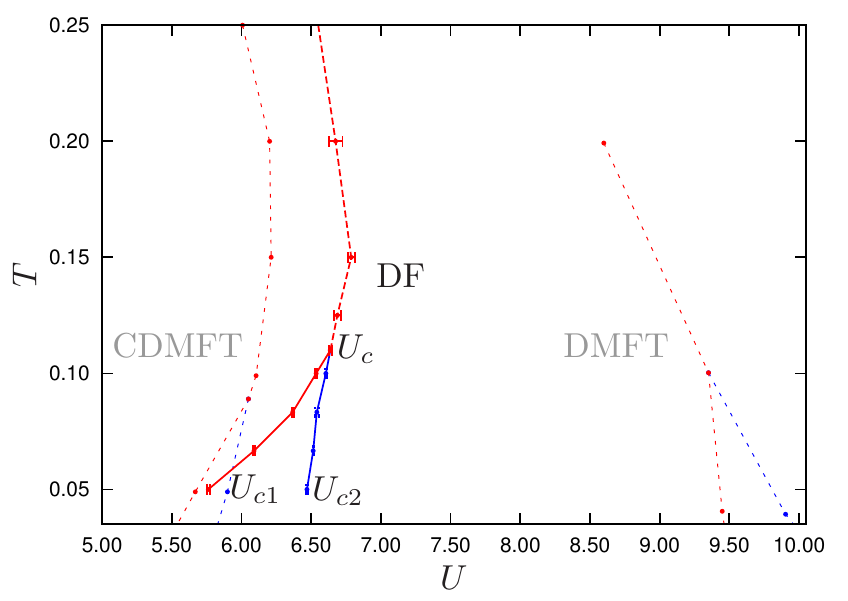} 
\caption{\label{fig:phasediagram} (Color online) Phase diagram of the two-dimensional Hubbard model obtained with dual fermion (DF) calculations. The critical point occurs at $U_{c}\sim 6.64$ and $T=0.11$. Corresponding data for cellular DMFT (CDMFT) and DMFT taken from Ref.~\onlinecite{Park08} are shown for comparison.
}
\end{figure}

\begin{figure*}[t]
\includegraphics[trim={0 1.95cm 0 0},clip]{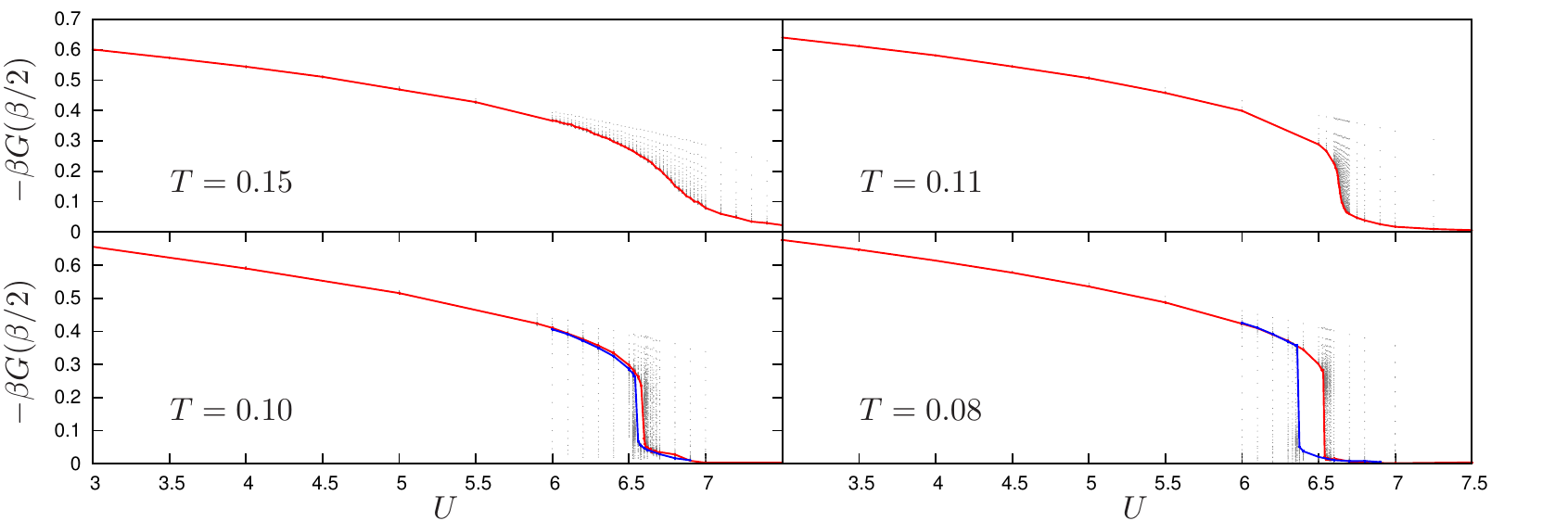}
\caption{\label{fig:convergence} (Color online) Estimator for the local density of states $-\beta G(\beta/2)\approx\pi A(0)$ as a function of $U$ for different temperatures. Results marked in red were obtained by starting the iterations with a metallic seed, while those marked by in blue were started from an insulating solution. Grey dots mark results from individual iterations and illustrate the convergence (for explanation see text).
}
\end{figure*}

We plot the phase diagram in the interaction-temperature plane of the two-dimensional Hubbard model at half filling in Fig.~\ref{fig:phasediagram}. 
The transition line $U_{c,1}$ (red in color) delimits the region of stability of the insulating phase, while $U_{c,2}$ delimits the stability region of the metallic phase.
We have determined the phase boundaries from the hysteresis of $\beta G(\beta/2)$, which approximates the density of states at the Fermi level. The transition lines encompass a coexistence region of metallic and insulating solutions. They merge at the critical point $U_{c}\sim 6.64$ at a temperature $T\sim 0.11$. Notably, their slope is positive. Above the critical point, we find a crossover region between the metallic and insulating states. The approximate position of the crossover line has been determined from the condition of maximum slope of $\beta G(\beta/2)$ vs $U$. 

DMFT and CDMFT results taken from Ref.~\onlinecite{Park08} are shown for comparison. In agreement with these methods, the Mott transition remains first-order in DF$^{(2)}$. In accordance with CDMFT, the critical interaction is significantly reduced compared to DMFT. Furthermore, both methods exhibit a positive slope of the transition lines which is reversed compared to DMFT. This is a consequence of the effects of spatial correlations and has previously been explained through an entropy argument~\cite{Park08}. In DMFT at $T=0$, the insulating solution represents a lattice of independent spin $s=1/2$ magnetic moments with a large residual entropy of $\log 2$ per site, while the entropy of the metal is lower. As temperature is increased, the insulator is preferred over the metal. In presence of spatial correlations, the residual entropy is greatly reduced and the situation is essentially opposite: the metal is preferred at elevated temperatures. As a result the slope of the transition lines is reversed.

The entropy also determines the slope of the crossover lines at high temperature. While the slope is positive immediately above the critical point, it becomes negative at very high temperature. In this regime, spatial correlations are thermally destroyed and the high-temperature insulating phase is favored because of its high entropy. The negative slope is therefore common among all three approaches.

The critical $U_{c}$ in DF$^{{(2)}}$ is substantially reduced compared to the DMFT value $U_{c}=9.35$~\cite{Park08} and in good agreement with $2\times 2$ CDMFT, which has the lowest value ($U_{c}=6.05$)~\cite{Park08} and even closer to the 16-site DCA value $U_{c}=6.53$~\cite{WernerPC13}.
It is likely that the four-site CMDFT cluster overestimates short-range correlations due to absence of correlations on intermediate length scales and therefore underestimates the critical $U$. Indeed the plaquette singlet ground state has the highest probability among the cluster eigenstates~\cite{Park08}. In the 16-site DCA calculation, we expect the nearest-neighbor correlations to be reduced, as the cluster size extends well beyond the plaquette. As a result the critical $U_{c}$ is higher compared to CDMFT. 
In the DF$^{(2)}$, correlations are included in principle up to the extension of the lattice and hence well beyond the DCA cluster. However, due to the truncation of the diagrammatic series, correlations are only treated approximately at any length scale. Since the dual Green's function decays fast in real space, correlations are effectively relatively short-ranged.
This may be the reason why the dual fermion result is even somewhat higher than in DCA. Nevertheless, the two values are remarkably close.

In Fig.~\ref{fig:convergence}, we plot $-\beta G(\tau=\beta/2)$, as a function of $U$ ($G$ denotes the local lattice Green's function) as an indicator of the local density of states $A(\omega=0)$ at the Fermi level and to distinguish metallic and insulating phases.
The phase diagram in Fig.~\ref{fig:phasediagram} has been determined from such curves. At high temperatures, we find a crossover from the metallic to the insulating state as the interaction is increased. As the temperature is lowered, the curves become very steep at the crossover and a coexistence region develops below the critical point (lower panels). The hysteresis is also visible in other observables, such as the impurity double occupancy (not shown).

In the same figure, we have included the values of $-\beta G(\beta/2)$ from individual dual fermion outer loop iterations, marked as gray dots. This provides information on convergence.
The uppermost dot for a given $U$ is the result from the first dual fermion iteration. It varies with $U$ because we start from a converged DMFT hybridization at the respective $U$ as the initial guess. 
In the lower panels, we additionally plot the individual values from iterations initialized with the same insulating seed.
In both cases, the final value is usually approached monotonously, although very close to the solution oscillations may occur. Points which are denser on a vertical scale indicate a decelerated convergence towards the final solution. One can clearly see this effect in the crossover regime at lower temperature (upper right panel), or at the edges of the hysteresis curves which delineate the coexistence region (lower panels).
The critical slowing down is also observed in DMFT. Overall the convergence properties are similar to DMFT.

\subsection{Momentum space differentiation}

\begin{figure}[t]
\includegraphics{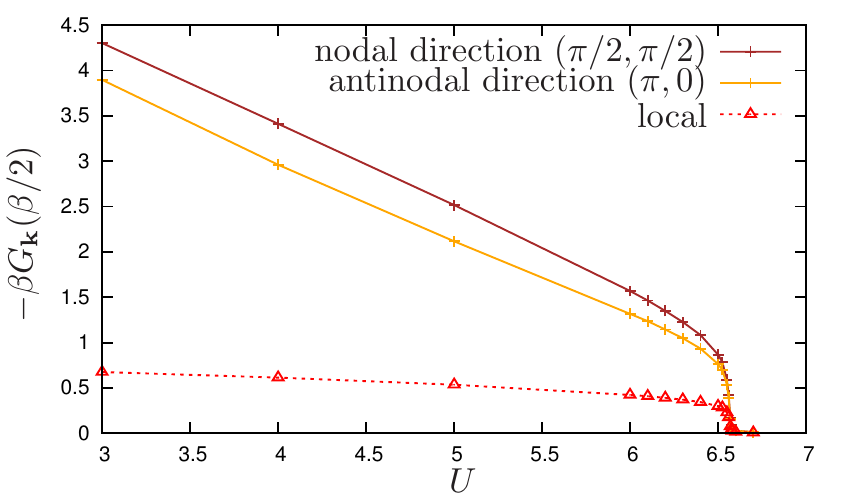} 
\caption{\label{fig:momdiffg} (Color online) Momentum differentiation in the dual fermion approach. The estimator $-\beta G_{\kv}(\beta/2)$ for the Fermi level spectral function is plotted for two k-points on the Fermi surface at $T=0.08$ as a function of $U$. The estimator for the local DOS is shown for comparison.
}
\end{figure}

\begin{figure}[t]
\includegraphics[width=\columnwidth]{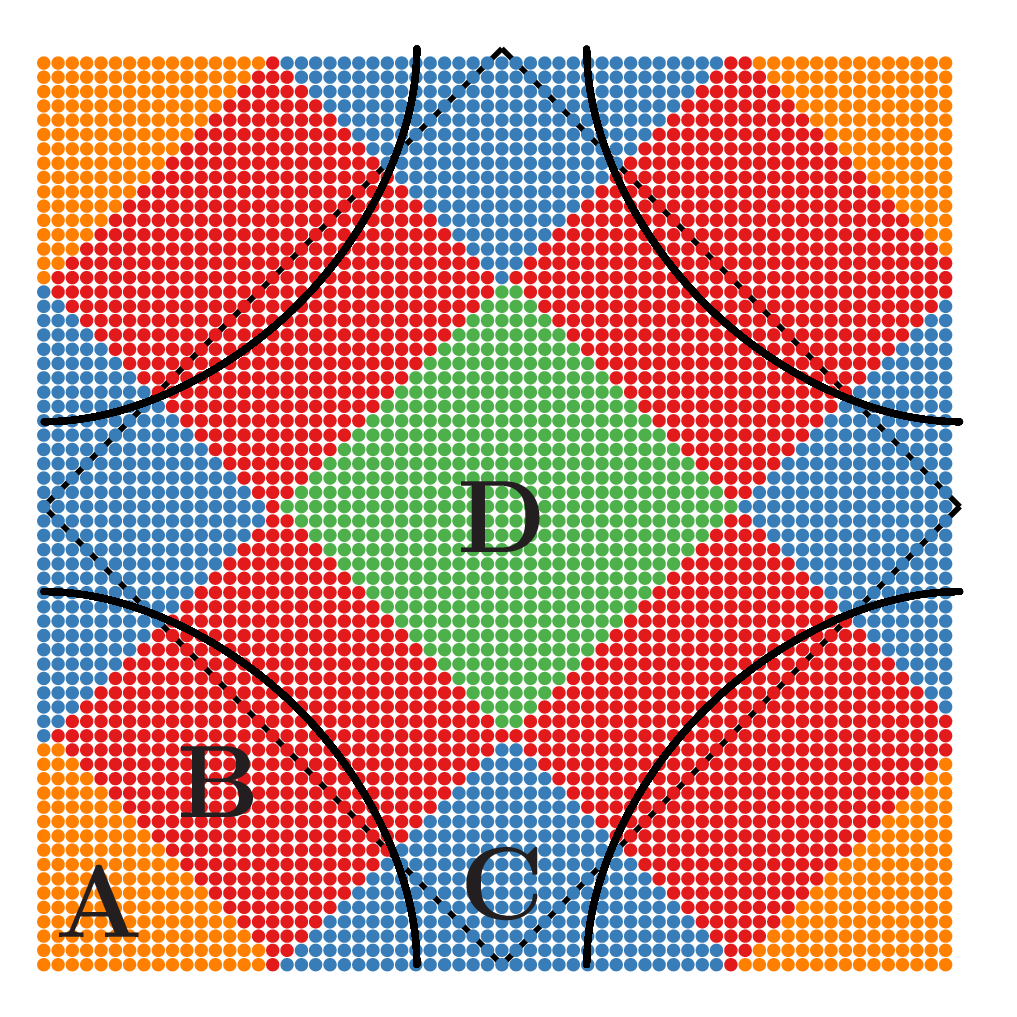} 
\caption{\label{fig:dcapatches} (Color online) Mapping of individual k-points in the dual fermion calculation to the corresponding DCA patches in an eight-site patch geometry. The number of points corresponds to the native momentum resolution used in the dual fermion calculations ($64\times 64$ in this case). 
The noninteracting Fermi surfaces for $t'/t=0$ (square) and $t'/t=-0.3$ are also shown. Patches B and C contain the Fermi surface. Patch B covers the nodal and patch C the antinodal direction.
}
\end{figure}

Because of the momentum dependence of the DF$^{(2)}$ self-energy one can expect some differentiation introduced through the diagrammatic corrections, which is absent in DMFT.
In Fig.~\ref{fig:momdiffg} we plot $-\beta G_{\kv}(\beta/2)$ for two high-symmetry points in the Brillouin zone, which correspond to the nodal $(\pi/2,\pi/2)$ and antinodal $(\pi,0)$ directions. 
A clear momentum space differentiation occurs sufficiently far from the transition. 
The value of the spectral function at the antinode is smaller than at the node, in accordance with the fact that quasiparticles are more strongly renormalized in this direction~\cite{Vilk97,Ferrero09,Werner09,Gull09,Rubtsov09}.
As the transition is  approached however, the differentiation diminishes and vanishes at the transition point.

This behavior is contrary to the DCA, where the transition manifests itself as orbital selective in momentum space: the self-energy at the antinode undergoes the transition before the one at the node.
To compare our results with those of the DCA, we coarse-grain the Green's function according to the DCA patches.

In DCA, the self-energy is a piecewise constant function in momentum space and is expressed in terms of basis functions $\phi_{\KV}(\kv)$ which are equal to one if $\kv$ is contained in the patch centered around $\KV$ and zero otherwise. Here we choose the eight-site patch geometry, because the nodal and antinodal directions can be well distinguished and sufficient data is available for comparison. In Fig.~\ref{fig:dcapatches} we show the mapping of the $64\times 64$ individual k-points of the DF calculation to the corresponding DCA patches. Each patch contains the same number of k-points. The $N_{p}=8$ patches can further be grouped by symmetry as indicated by color. The patches of type B contain the nodal and the type C patches the antinodal points. Since the Brillouin zone is tiled completely by all patches, they have the property that
\begin{align}
\sum_{\KV=1}^{N_{p}}\phi_{\KV}(\kv)=1.
\end{align}
The \emph{coarse-grained} Green's function is defined as
\begin{align}
\bar{G}_{\KV}(\inu) = \frac{N_{p}}{N}\sum_{\kv}\phi_{\KV}(\kv)G_{\kv}(\inu),
\end{align}
so that the local Green's function is obtained as a momentum integral over the patch momenta,
\begin{align}
G(\inu) = \frac{1}{N_{p}}\sum_{\KV}\bar{G}_{\KV}(\inu) = \frac{1}{N}\sum_{\kv}G_{\kv}(\inu).
\end{align}

\begin{figure}[t]
\centering
\includegraphics{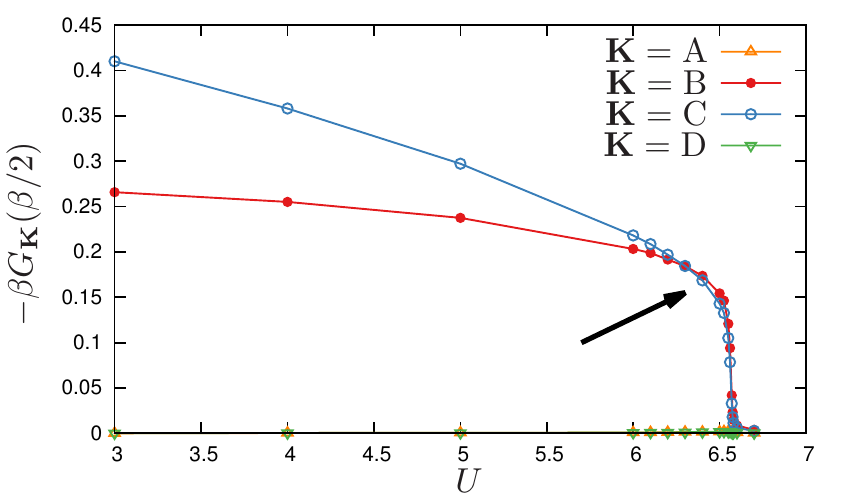} 
\caption{\label{fig:momdiffgpatch} (Color online) Estimator for the density of states  integrated over the momentum patches of Fig.~\ref{fig:dcapatches} as a function of $U/t$ approaching the Mott transition in DF$^{(2)}$ for fixed temperature $T/t=1/12$ and $t'=0$.
The plot shows $-\beta G_{\KV}(\beta/2)$ rescaled by the number $n_{\alpha}$ of equivalent patches, where $n_{\text{A}}=1$, $n_{\text{B}}=4$, $n_{\text{C}}=2$ and $n_{\text{D}}=1$. The Fermi surface is entirely contained in patches B and C, so that the contribution on patches A and D is negligibly small.
}
\end{figure}

\begin{figure}[t]
\centering
\includegraphics{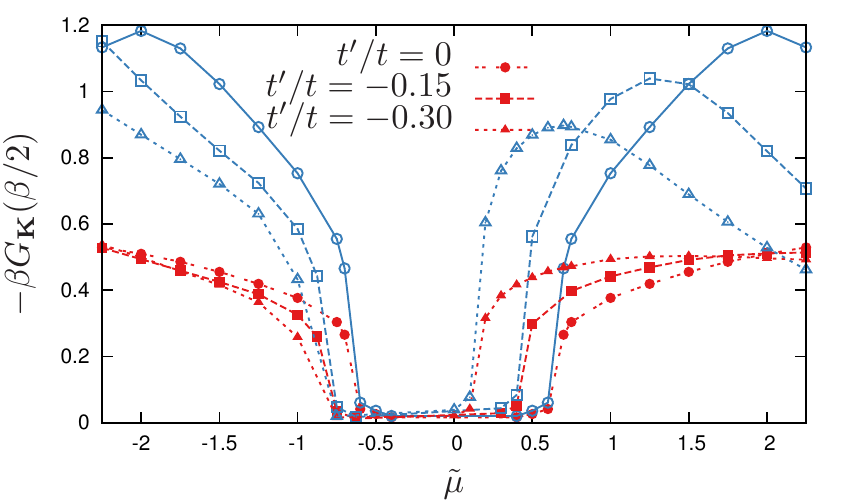} 
\caption{\label{fig:dopingdrpatch} (Color online) Quantity $-\beta G_{\KV}(\beta/2)$ as in Fig.~\ref{fig:momdiffgpatch} but for the doping driven transition as a function of $\tilde{\mu} = \mu-U/2$ for $U/t=7$, $T/t=0.08$ and different values of the next-earest-neighbor hopping $t'$ on the patches $A$ and $B$. The color coding corresponds to the one in Fig.~\ref{fig:dcapatches}.
}
\end{figure}

We plot $-\beta G_{\KV}(\beta/2)$ obtained from the coarse-grained Green's function on the four different patches in Fig.~\ref{fig:momdiffgpatch}. We observe that sufficiently below the transition, the result in sector B (around the nodal point) is smaller than that in sector C, which contains the antinodal point~\footnote{This can be understood from the non-interacting band structure. The antinodal point is a Van Hove singularity at the Fermi energy, i.e., $E_\kv=0$ and $\nabla_\kv E_\kv =0$, so that there are many states close to the Fermi energy in sector C.}.
This seems contrary to the previous result of Fig.~\ref{fig:momdiffg}. Close to the transition however, the sector C value shifts below the sector B value as indicated by the arrow. It remains smaller all the way to the transition. This is in qualitative agreement with the DCA results of Fig.~8 of Ref.~\onlinecite{Gull09}. What we do not observe, however, is a gap opening in sector C before it happens in sector B. In second-order DF, the gap appears to open in both sectors simultaneously. The difference between the two sectors is also much smaller.
Taken together with the results of section~\ref{sec:Mott}, we find that while our method and cluster methods agree qualitatively concerning the strong reduction of the critical interaction and the inversion of the slope of the transition lines, the momentum differentiation is much smaller than predicted by the DCA.

%This behavior actually looks similar to the frustrated case (finite $t'$) in DCA~\cite{Gull09}. We speculate that singlet correlations between these patches, included in the DCA but absent in the present approach, are responsible for the orbital selective character of the transition. They are weakened through frustration, so that our results at least qualitatively resemble the DCA ones for finite $t'$.

In Fig.~\ref{fig:dopingdrpatch} we show results similar to the ones in Fig.~\ref{fig:momdiffgpatch}, albeit for the doping driven transition for different values of the next-nearest-neighbor hopping $t'$. Similarly to the case before we find that the momentum space differentiation vanishes at the transition.

\subsection{Higher-order diagrams}

\begin{figure}[h]
\includegraphics{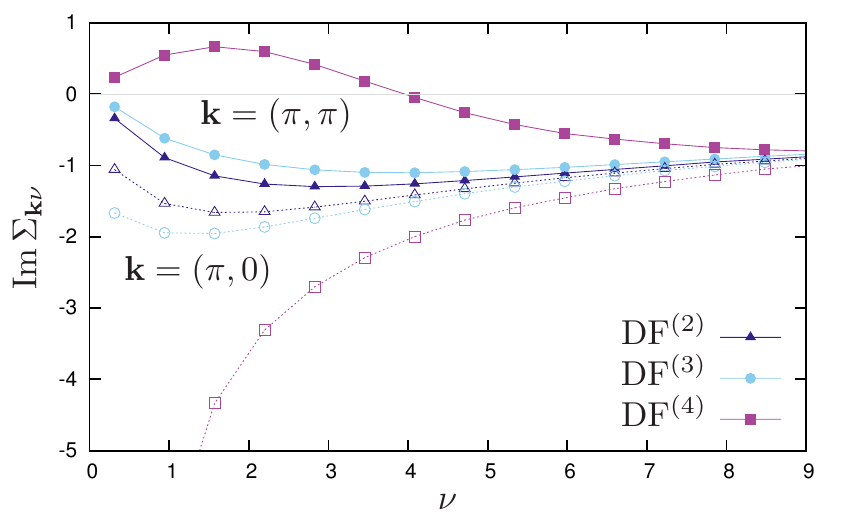} 
\caption{(Color online) Comparison of the self-energy at $U/t=6.2$ and $T/t=0.1$ between DF$^{(2)}$ with higher-order dual approximations. The solid figures correspond to $\kv=(\pi,\pi)$, the empty symbols to $\kv=(\pi,0)$.}
\label{fig:DFn}
\end{figure}

The results presented so far were all been obtained using the DF$^{(2)}$ approximation, i.e., using \eqref{sigmad2} for the self-energy. It is possible to take into account higher-order diagrams in the dual theory. Two common techniques for this are the ladder approximation~\cite{Hafermann09} and diagrammatic Monte Carlo techniques~\cite{Iskakov2016,Gukelberger2017}. 
Here we compare DF$^{(2)}$ with higher-order DF approaches: DF$^{(3)}$ and DF$^{(4)}$. That is, we add the third- and fourth-order ladder diagrams to the expression for the dual self-energy in Eq.~\eqref{sigmad2}.

Fig.~\ref{fig:DFn} shows the physical self-energy at $U/t=6.2$, $T/t =0.1$. In DF$^{(2)}$, these parameters are close to the transition and the convergence of the outer loop is relatively slow, c.f. Fig.~\ref{fig:convergence}. Figure~\ref{fig:DFn} shows that the inclusion of higher-order diagrams leads to  more momentum differentiation compared to DF$^{(2)}$. However, we also observe that the self-energy in DF$^{(4)}$ actually has an unphysical positive imaginary part.

These results indicate that the DF series behaves as an asymptotic series, in which the lowest-order approximation yields reasonable results in accordance with other methods, whereas higher-order approximations of a divergent series may become unphysical.  Consistent with Figure 10 of Ref.~\cite{Gukelberger2017}, these parameters are deep inside the region inside the regime of poor convergence of the series sampled using a diagrammatic Monte Carlo approach. 
This should be kept in mind if the series is outside of its convergence radius. 
Convergence of the series can be easily tested by evaluating the leading eigenvalue of the matrix $(-T/N)\sum_{\nu'\kv'}\gamma_{\nu\nu'\omega}\tilde{G}_{\kv\nu'}\tilde{G}_{\kv+\qv \nu'+\omega}$.
The divergence is closely related to the perfect nesting of the Fermi surface and indicative of a (spurious) second-order transition to the antiferromagnetic state because of the mean-field starting point. Away from half-filling or on frustrated lattices, this problem is less of an issue. On the triangular lattice, the second-order approximation gives results which are very close to 16-site DCA results~\cite{Lee2008}.
We emphasize that the infinite ladder approximation does not suffer from this shortcoming. In the latter, convergence of an initially diverging series can be ensured through a self-consistent renormalization of the dual Greens functions~\cite{Otsuki2014}. It however describes different physics, as it includes long-range correlations on all length scales.

\section{Self-consistency conditions}
\label{sec:sc}

\begin{figure}[t]
\includegraphics{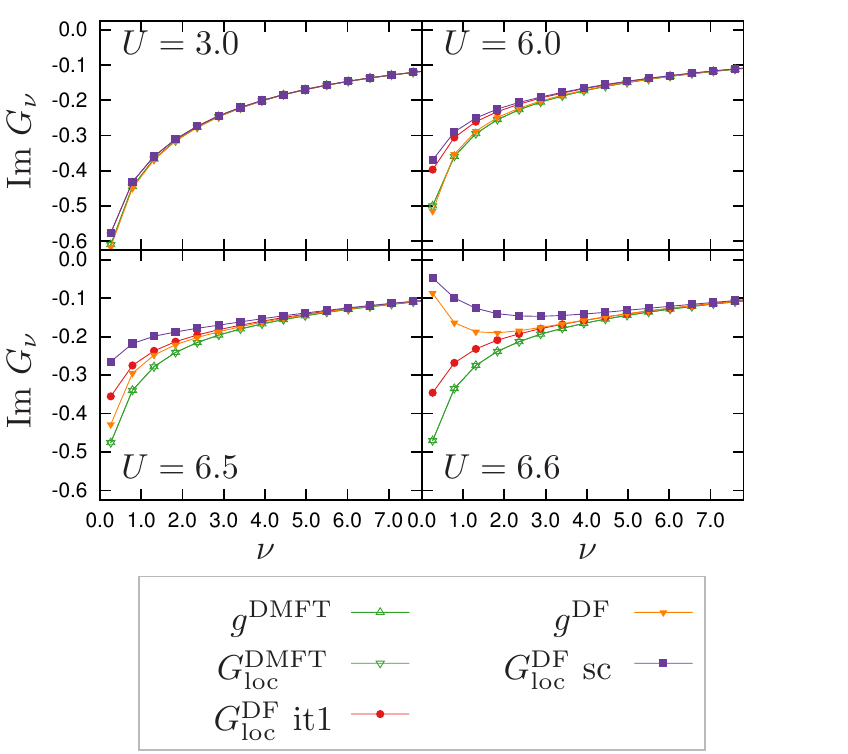}
\caption{\label{fig:gloc} (Color online) Comparison of impurity and local lattice Green's functions in DMFT and DF with the dual self-consistency condition (i), at $\beta t=12$. In contrast to DMFT, in DF the impurity model Green's function does not equal the local lattice Green's function. 
}
\end{figure}

Previous DF results were exclusively based on the dual self-consistency condition (i), Eq.~\eqref{eq:sc1}.
In this section we compare the dual and lattice self-consistency conditions.

The outer self-consistency modifies the auxiliary quantum impurity model, with the idea that it captures a much of the correlation effects as possible. The importance of this outer self-consistency was already illustrated in Fig.~\ref{fig:convergence}. As discussed in Sec.~\ref{sec:method}, it is possible to contemplate alternative self-consistency conditions, as was also done in other DMFT-based methods~\cite{Stepanov16,vanLoon16}.

We compare the effect of the two DF self-consistency conditions based on the impurity and local lattice Green's function in DMFT and DF.
Fig.~\ref{fig:gloc} shows the result for the condition (i), Eq.~\eqref{eq:sc1}. In all four panels the DMFT impurity Green's function is equal to the local lattice DMFT Green's function, $g^{\text{DMFT}}=G_{\text{loc}}^{DMFT}$, as required by DMFT self-consistency.
In DF however, the impurity Green's function ($g^{DF}$) and the local lattice Green's function ($G^{\text{DF}}_{\text{loc}}$) are different.
For small $U$ (top left panel) the DF impurity Green's function $g^{\text{DF}}$ remains essentially unchanged during the outer loop self-consistency. However both show a visible correction compared to the DMFT result, which corresponds to a suppression of spectral weight at the Fermi level.
For larger values of $U$ closer to the transition this correction increases and also the impurity Green's function in DF differs substantially from DMFT.

The lower right panel further shows that the outer loop iterations are essential to capture the Mott transition: the result from the first DF iteration is based on a metallic environment where the hybridization is equal to its DMFT value. $G_{\text{loc}}^{\text{DF}}$ corresponds to a diagrammatic expansion around DMFT and yields a metallic solution.
The self-consistent solution $G_{\text{loc}}^{\text{DF}}$ sc however is insulating, as is the impurity model. 

An explanation may be that the Mott transition is non-perturbative and hence cannot be reached by expanding around a metallic state. On the other hand, the condition (i) corresponds to the summation of an infinite partial series: all diagrams with a local loop are absorbed into the impurity model.
Furthermore, the perturbation series contains non-local contributions to the self-energy,  which give rise to local contributions to the dual Green's function. Non-local correlations therefore effectively alter the local environment (the hybridization function).

\begin{figure}[t]
\includegraphics{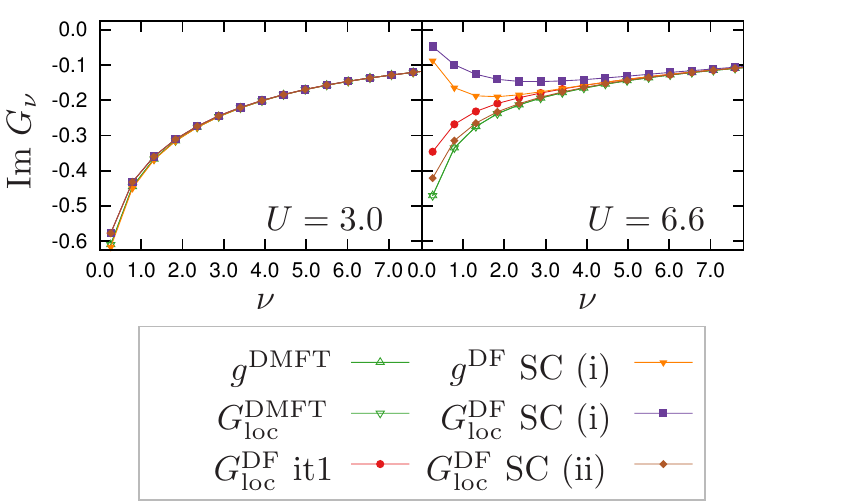}
\caption{\label{fig:alternative_sc} (Color online) Comparison of impurity and local lattice Green's functions in DMFT and DF using two different self-consistency conditions, Eqs.~\eqref{eq:sc1} and \eqref{eq:sc2}. This corresponds to the top left and bottom right panels in Fig.~\ref{fig:gloc}, with the additional brown diamons denoting the results of self-consistency condition (ii), i.e., Eq.~\eqref{eq:sc2}. Note that $g^{\text{DF}}$ is equal to $G^{\text{DF}}_{\text{loc}}$ when using selfconsistency condition (ii), so that the former is not shown in the Figure. 
}
\end{figure}

The interpretation of condition (ii) in terms of dual diagrams is not obvious. Fig.~\ref{fig:alternative_sc} compares results for the two conditions.
For small $U$ (left panel) well below the transition, the impurity model remains essentially the same as in DMFT using either self-consistency condition. However, for large $U$ (right panel), the results of the two self-consistency conditions are notably different. The dual self-consistency condition (i) results in an insulating solution, while the lattice condition (ii) actually mediates between the DMFT solution and the first dual iteration and does not capture the transition to the insulating state. 
We have also performed calculations using the lattice self-consistency condition at larger interaction strengths, above the DMFT MIT, and did not find an insulating state. However, above the DMFT MIT, the simulations with this self-consistency condition were hard to stabilize due to issues that will be discussed below.

\subsection{Self-consistency away from half-filling}

Additional understanding of the role of the self-consistency condition can be gained by moving away from half-filling. The motivation for the traditional DF self-consistency condition \eqref{eq:sc1} is that it ensures that Hartree-like diagrams (the first diagram in Fig.~\ref{fig:2ndorder}) vanish~\cite{Rubtsov08}. Looking at an asymptotic expansion of $\Sigma_{\kv\nu}$ in terms of $\nu$, this means that there are no non-local contributions to the constant and $(i\nu)^{-1}$ terms and the leading dual contributions occur at order $(i\nu)^{-2}$. 
This is of practical importance since the dual contribution requires knowledge of the vertex $\gamma_{\nu\nu'\omega}$, and the fast decay means that the vertex only needs to be determined for a limited number of fermionic frequencies. Apart from this practical computational issue, the vanishing Hartree diagram also has important theoretical considerations. 

A point of attention for diagrammatic extensions based on impurity models is that the final local observables of the theory are generally not identical to those of the impurity model~\cite{vanLoon16,Rohringer16,Gukelberger2017,Krien17}. 
In particular, away from half-filling the density can be inconsistent between the impurity model and the lattice theory~\footnote{The case of half-filling in the square lattice Hubbard model is rather special. There is particle-hole symmetry, the vertex $\gamma_{\nu\nu'\omega}$ is purely real and the Green's function $\tilde{G}_{\nu}$ is purely imaginary. This combination ensures that there is no contribution to the real part of the self-energy from the Hartree diagram. This explains why the inconsistencies discussed below are less important at half-filling: The particle-hole symmetry makes the leading order inconsistency go away. Above, we mentioned that it was hard to converge calculations using the lattice self-consistency condition \eqref{eq:sc2} for $U>U^{\text{DMFT}}_{\text{MIT}}$. The origin of this instability lies in the density which always deviates from half-filling due to Monte Carlo errors, in this regime these deviations are unstable within the self-consistent loop and led to larger and larger deviations of the density from half-filling.}. 
As explained above, using self-consistency condition \eqref{eq:sc1} ensures that the Hartree diagram vanishes so that 
\begin{align}
\lim_{\nu\rightarrow \infty} \Re \Sigma_{\kv\nu} \overset{\text{SC (i)}}{=} \lim_{\nu\rightarrow \infty} \Re \Sigma_{\text{imp}} = \frac{U}{2} \av{n}_{\text{imp}}. \label{n1:sc1}
\end{align}
In this sense, the self-energy with non-local corrections is consistent with the density of the impurity model. The same holds for the $(i\nu)^{-1}$ term.
On the other hand, there is a mismatch between the single-particle Green's function of the lattice and that of the auxiliary impurity model,
\begin{align}
\Tr G \overset{\text{SC (i)}}{\neq} \Tr g = \av{n}_{\text{imp}}. \label{n2:sc1}
\end{align}
Combining Eqs.~\eqref{n1:sc1} and \eqref{n2:sc1} shows that the dual approximation does not satisfy this property of the exact solution
\begin{align}
 \frac{U}{2} \Tr G \overset{\text{SC (i)}}{\neq} \lim_{\nu\rightarrow \infty} \Re \Sigma_{\kv\nu}.
\end{align}

The alternative self-consistency condition \eqref{eq:sc2} enforces $G^\text{loc}=g$ and thus also
\begin{align}
\Tr G \overset{\text{SC (ii)}}{=} \Tr g = \av{n}_{\text{imp}}. \label{n2:sc2}
\end{align}
This comes at a cost, since there is now a finite dual contribution to the asymptotics of the self-energy and 
\begin{align}
\lim_{\nu\rightarrow \infty} \Re \Sigma_{\kv\nu} \overset{\text{SC (ii)}}{\neq} \lim_{\nu\rightarrow \infty} \Re \Sigma_{\text{imp}} = \frac{U}{2} \av{n}_{\text{imp}}, \label{n1:sc2}
\end{align}
which again implies
\begin{align}
 \frac{U}{2} \Tr G \overset{\text{SC (ii)}}{\neq} \lim_{\nu\rightarrow \infty} \Re \Sigma_{\kv\nu}.
\end{align}

We should stress that these arguments hold when only diagrams containing two-particle vertices are used. Diagrams with higher-order vertices do contribute to the local asymptotics. The dual transformation is an exact transformation, so that the exact solution of the dual action (all diagrams with all vertices taken into account) results in the exact solution of the original model and in particular satisfies all analytical properties of the exact solution.

\begin{figure}[t]
\includegraphics{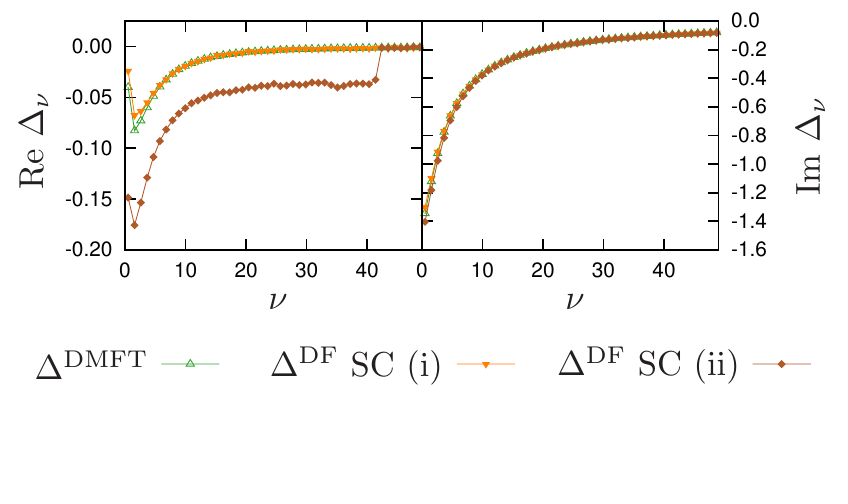}
\caption{\label{fig:delta_SC} (Color online) The hybridization function at $U/t=4$, $\mu/t=1$, $\beta t=6$, in DMFT and in self-consistent DF using self-consistency conditions \eqref{eq:sc1} and \eqref{eq:sc2}. The cut-off for dual self-energy corrections is at $\nu_{\text{max}}=(2\cdot 40+1)\pi T \approx 42$.
% DMFT 0.8208
% DF sc2, imp   0.819
% DF sc2, G     0.84
% DF sc1, imp   0.836
% DF sc1, sigma 0.81
}
\end{figure}

From a practical point of view, \eqref{n1:sc2} is problematic for the frequency cut-off in the dual calculations. It essentially means that any finite frequency cut-off $\nu_\text{max}$ in the calculation of $\tilde{\Sigma} $ always leads to a \emph{finite} discontinuity in $\Re \Sigma$ at $\nu_\text{max}$. This differs from the traditional self-consistency condition, where the magnitude of the discontinuity in $\Sigma$ decays as $(i\nu)^{-2}$. To illustrate this, Fig.~\ref{fig:delta_SC} shows the self-consistently determined hybridization function away from half-filling. In the results using \eqref{eq:sc2} (brown diamonds), the real part has a discontinuity at the frequency corresponding to the cut-off used for dual diagrams, even though this cut-off happens at a rather large energy, approximately 5 times the bandwidth. Thus for calculations away from half-filling based on the self-consistency condition \eqref{eq:sc2} is impractical unless measures are taken to ensure the proper frequency asymptotics of the dual diagrams.

\section{Conclusions}

We have investigated the second-order dual fermion approximation and how it captures the Mott transition in the two-dimensional Hubbard model: The transition is first order and the critical interaction is strongly reduced compared to (single-site) DMFT. The results are in close agreement with results obtained from cluster approaches. The approximation also captures the inversion of the slope of the transition lines with respect to DMFT. We interpret this as an effect of the dynamical short-range nearest-neighbor antiferromagnetic correlations which are incorporated through the diagrams and which reduce the entropy of the paramagnetic metallic state.

While the strong reduction of the critical interaction and the inversion of the slope of the transition lines are in qualitative agreement with cluster DMFT results, we find a much weaker sector selective momentum differentiation. Contrary to DCA, it disappears at the Mott transition. 
It would be interesting to see whether the sector-selective character of the DCA solution is increased or decreased when augmented with DF corrections in a  cluster dual fermion formulation~\cite{Hafermann08,Yang11,Iskakov18}.

The second-order approximation is a relatively simple approximation which allows one to study nontrivial effects of short-range dynamical spatial correlations in correlated systems beyond DMFT.
Including truly long-ranged correlations that respect the Mermin-Wagner theorem requires at least a ladder approximation.

We found that the dual perturbation series behaves like an asymptotic series. Higher-order diagrams yield unphysical results at low temperature, so that the approximation should be used with care outside its radius of convergence. This is in line with the observation that the DF diagrammatic series may converge to wrong results at low temperature~\cite{Gukelberger2017}. We note that the ladder approximation, on the contrary, yields sensible results~\cite{Hafermann09,Gukelberger2017,vanLoon17}.

We further reported results based on the alternative self-consistency condition akin to the one used in DMFT. This condition does not capture the Mott transition. It also changes the asymptotics of the physical fermion self-energy, making it less relevant in practice.
This is consistent with the recent findings of Ref.~\onlinecite{Ribic18}. 
The conventionally used self-consistency condition, on the other hand, corresponds to the elimination of an infinite partial series from the diagrammatic expansion, which may be crucial to obtain an insulating solution based on the metallic DMFT starting point.

\begin{acknowledgments}
The simulations employed a modified version of an open source implementation of the hybridization expansion quantum impurity solver\cite{Hafermann13}, based on the ALPS libraries\cite{ALPS2}.
The authors would like to thank Tin Ribic for useful discussion.
E.G.C.P. v. L. and M.I.K. acknowledge support from ERC Advanced Grant 338957 FEMTO/NANO.
\end{acknowledgments}

\appendix

\bibliography{main}

\end{document}